\documentstyle[psfig,conf_iap,]{article}
\begin{document}
\heading{%
%
Accelerating Universe Without Event Horizon
%
}
\par\medskip\noindent
\author{%
Pedro F. Gonz\'{a}lez-D\'{\i}az$^{1}$
}
\address{%
Instituto de Matem\'{a}ticas y F\'{\i}sica Fundamental, CSIC,\\ Serrano
121, 28006 Madrid, Spain.
}

\begin{abstract}
It is shown that if a small negative cosmological constant is
added to quintessence models with equation of state $p=\omega\rho$
on the range $-1 <\omega < -1/3$, the resulting scenarios could
not contain any future event horizons. Therefore, such
cosmological accelerating scenarios could not present the kind of
obstacles recently pointed out to define a set of observable
quantities analogous to an S-matrix needed by predictive M and
string theories in an accelerating universe driven by dark energy.
These scenarios become accelerating only at early times after
coincidence time, and lead finally to re-collapse.
\end{abstract}
\vspace{.5cm}

Perhaps one of the biggest problems of nowadays physics is the
possible incompatibility between the recently discovered
accelerated expansion of the universe and a consistent
mathematical formulation of string or M theories based on the
existence of a S-matrix or S-vector, provided the universe would
accelerate eternally due to the existence of a positive
cosmological constant or quintessence scalar field with constant
parameter for its equation of state, and hence shows a future
event horizon \cite{FKMP},\cite{HKS}. Solutions to this problem
have been considered only at the cost of allowing loss of quantum
coherence \cite{EMN} or the existence of real light signals
traveling backward in time \cite{Gon1}. In this report we present
a very simple solution to the problem which is based on the
simultaneous existence of both a quintessential slowly-varying
scalar field $\phi$ and a negative cosmological constant $\Lambda$
\cite{CGOQ} whose absolute value is very small, so that the
deceleration parameter keeps negative along the entire evolution
after coincidence time. Disregarding the effects of any matter
field (see later on), let us then use the action
\begin{equation}
S=\int d^4 x\sqrt{-g}\left[\frac{1}{16\pi
G}\left(R-2\Lambda\right) +L_{\phi}\right] ,
\end{equation}
where all symbols have their conventional meaning, and $L_{\phi}$
is the Lagrangian for the quintessence field,
$L_{\phi}=-\phi_{,n}\phi^{,n}-V(\phi)$, $V(\phi)$ being the
quintessence potential whose explicit form is not of interest for
our present purposes \cite{Gon}. In the most observationally
favored case of a spatially flat universe \cite{Deb}, if we use
the customary definition of the quintessence field, $8\pi
G\rho/p=\dot{\phi}^2+/-V(\phi)$ (with $\dot{x}=dx/dt$), a state
equation $p=\omega\rho$, with $-1/3 <\omega < -1$ being constant,
and the conservation law $\rho_{\phi}\propto a^{-3(1+\omega)}$
\cite{Gon}, then we obtain the Friedmann equations for the scale
factor $a$
\begin{equation}
\dot{a}^2 =\frac{8\pi G}{3\alpha}a^{-(1+3\omega)} +\lambda a^2
\end{equation}
\begin{equation}
\ddot{a}=-\frac{4\pi
G}{3\alpha}(1+3\alpha)a^{-(2+3\omega)}+\lambda a,
\end{equation}
where $\alpha$ is an integration constant and
$\lambda=\Lambda/(3H_0^2)$, with $H_0$ is the current value of the
Hubble parameter. We have been able to obtain the general solution
to these differential equations, that is
\begin{equation}
a(t)= \left\{\sqrt{\frac{8\pi G}{3\alpha\lambda}}
\sinh\left[\frac{3(1+
\omega)\sqrt{\lambda}t}{2}\right]\right\}^{2/[3(1+\omega)]} .
\end{equation}
Note that this describes an eternally rapidly accelerating flat
universe and becomes the solution for a constant quintessence
field used by Fischler {\it et al.} \cite{FKMP} and Hellerman {\it
et al.} \cite{HKS} in the limit $\sqrt{\lambda}t/2 <<1$. By the
same reasons as given by these authors, it is again deduced that
also for solution (4) there will be a future event horizon which
conflicts the formulation of any fundamental theory based on the
existence of an S-matrix at any infinite distances. However, if we
change sign of $\Lambda$, $\lambda\rightarrow -\lambda$, and use a
small enough value for $|\lambda|$ such that the deceleration
parameter \cite{Gon} $q_0=\left[\Omega_M+
\Omega_{\phi}(1+3\omega)+|\lambda|\right]/2$ (where the $\Omega$'s
are the usual dimensionless cosmological parameters for matter
($M$) and the quintessence field ($\phi$)) keeps a negative value
for sufficiently large $|\omega|$, then the resulting general
solution
\begin{equation}
a(t)= \left\{\sqrt{\frac{8\pi G}{3\alpha |\lambda|}}
\sin\left[\frac{3(1+
\omega)\sqrt{|\lambda|}t}{2}\right]\right\}^{2/[3(1+\omega)]},
\end{equation}
will not show any future event horizon, while keeping the same
dynamical accelerating behaviour at small values of
$3(1+\omega)\sqrt{|\lambda|}t/2$ as the solution used in Refs.
\cite{FKMP} and \cite{HKS}. Assuming then that we are now living
in such a accelerating regime, one should expect that, instead of
an event horizon, the future of our universe will show a
re-collapsing behaviour tending the size of the universe to
recover that at the coincidence time. In order to see this in more
detail, let us consider an equation of state for $\phi$ with
$\omega=-2/3$. In that case, the flat FRW metric reduces to
\begin{equation}
ds^2= -dt^2 +\left(\frac{8\pi G}{3\alpha|\lambda|}\right)^2
\sin^4\left(\frac{\sqrt{|\lambda|}t}{2}\right)\left(dr^2+r^2
d\Omega_2^2\right) ,
\end{equation}
where, by taking the initial condition $a(t)\rightarrow 0$ as
$t\rightarrow 0$, $0 \leq t\leq +2\pi$, $r\geq 0$ is the radial
coordinate and $d\Omega_2^2$ is the metric on unit two-sphere.
Following next the procedure used by Hawking and Ellis \cite{HE},
metric (6) can be transformed into a line element that is defined
in terms of a conformal time
\begin{equation}
\eta-\eta_0 =-\frac{3\alpha\sqrt{|\lambda|}}{4\pi
G}\tan^{-1}\left(\frac{\sqrt{|\lambda}t}{2}\right),
\end{equation}
in which $\eta_0$ is an integration constant, $-\infty <\eta
<+\infty$, and (redefining $\eta=\eta-\eta_0$)
\begin{equation}
a(\eta)=\frac{8\pi G}{3\alpha|\lambda|}\left[1+\left(\frac{4\pi
G}{3\alpha}\right)^2\frac{\eta^2}{|\lambda|}\right]^{-1} ,
\end{equation}
which is conformal to that part of the Einstein static universe
defined by the interval covered by time $\eta$, $-\infty <\eta
<+\infty$ (see Fig. 1(i)), with a conformal factor
$\frac{1}{4}a(\eta)^2 \sec^2\left[\frac{1}{2}(t'+r')\right]
\sec^2\left[\frac{1}{2}(t'-r')\right]$, in which the new
coordinates $r'$ and $t'$ are related with $r$ and $\eta$ by the
expression provided on page 121 of Ref. \cite{HE}. Metric (6) has
an apparent singularity at $t=0$ (i.e. $\eta=-\infty$), but does
not show any event horizon.

\begin{figure}
\centerline{\vbox{
\psfig{figure=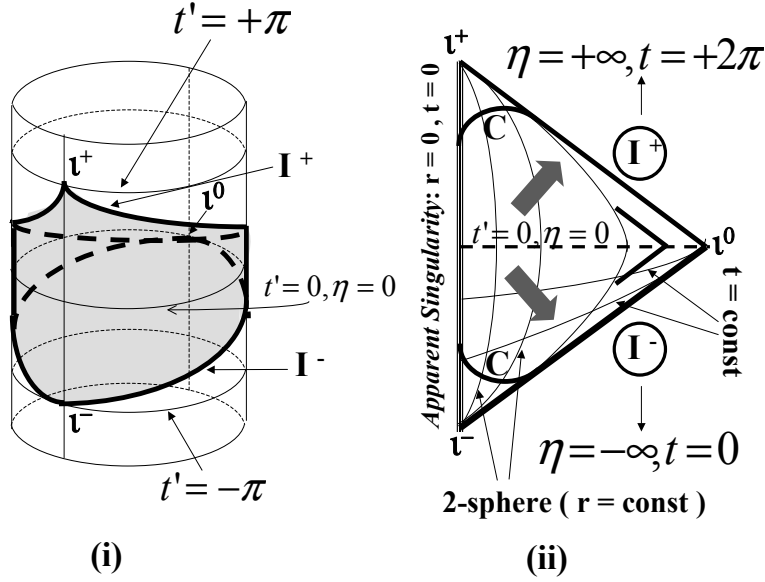,height=10.cm} }} \caption{(i)
Region of the Einstein static universe (represented by an embedded
cylinder) which is conformal to the whole of the space-time
described by metric (6). (ii) The Penrose-Carter diagram of the
space-time described by metric (6). Each point, except $\iota^0$,
$\iota^+$, $\iota^-$ and those on $r=0$, represents a two-sphere.
The corrections denoted as C on the diagram express the effects of
introducing a radiation field. }
\end{figure}

The latter conclusion can most easily be seen by checking that the
quantity $\bigtriangleup=|\eta-\eta_0|_{t=t_0}^{t=2\pi}$ diverges
for any finite $t_0$ \cite{FKMP}, leaving no inaccessible region
for an observer on $r=0$. Moreover, a possible apparent horizon
must generally satisfy \cite{Bou}, \cite{HKS}
$\eta=2r/(1+3\omega)$, so $r=-\eta/2$ for $\omega=-2/3$, and hence
$\eta=2\tan[(t'+r')/2]=-2r$ and
$t'-r'=\tan^{-1}\{\tan[(t'+r')/2]-2r\}$. We can readily see that
no horizon can exist satisfying these expressions and $r\geq 0$
(i.e. $r'\geq 0$) simultaneously. Now, since our metric is
conformal to Einstein static universe on the region showed in Fig.
1(i), the Penrose-Carter diagram is as given in Fig. 1(ii).
Following the general procedure put forward by Bousso \cite{Bou},
we have also determined the holographic screens to be placed at
the two infinities at $\eta=\pm\infty$. Nevertheless, the whole
cosmological model we have considered so far does not match any
kinds of the realistic standard models. Having nearly a seventy
percent of dark energy \cite{Per} in the form of a quintessence
scalar field plus a negative cosmological constant does not
prevent taking into account the effects produced by the
cosmological parameter $\Omega_M$. Notwithstanding, for the
purposes of the present report it will suffice to add a radiation
field (e.g. a massless scalar field $\varphi$ which is conformally
coupled to gravity), amounting to an term $-4\pi G\varphi^2/3
-\varphi_{,n}\varphi^{,n}$, to the quantities in squared brackets
of action (1). After performing the calculation with that
additional field, we obtain that for early times the scale factor
goes as $a\propto t^{1/2}$ and as the solution (4) at later times.
This gives rise to the slight modifications of the conformal
diagram shown in Fig. 2, which do not lead to any change in the
main conclusions obtained before. Thus, the adding of a small
negative cosmological constant to the quintessence models makes
the resulting accelerating expansion of the universe to be non
eternal while preventing the eventual emergence of any event
horizon in its future. The mathematical formulation of M and
string theories seems to become therefore safe from all
cosmological challenges originally posed by dark energy.

\acknowledgements{The author thanks Carmen L. Sig\"{u}enza for useful
discussions. This work was supported by MCYT under Research
Project No. BMF2002-03758.}

\vfill

\begin{iapbib}{99}{
\bibitem{Bou} Bousso R., 1999, JHEP 9907; 1999, JHEP 9906; 2000,
class. Quant. Grav. 17, 997
\bibitem{CGOQ} C\'{a}rdenas R., Gonz\'{a}lez T., Mart\'{\i}n O., Quir\'{o}s I.,
astro-ph/0209524
\bibitem{Deb} DeBernardis P., {\it et al.}, 2000, Nature 404, 955
\bibitem{EMN} Ellis J., Navromatos N.E., Nanopoulos D.V.,
hep-th/0105206
\bibitem{FKMP} Fischler W., Kashani-Poor A., McNees R., Paban S.,
2001, JHEP 0107
\bibitem{Gon1} Gonz\'{a}lez-D\'{\i}az P.F., 2001, Phys. Lett. B522, 211;
2002, Phys. Rev. D65, 104035; Chernin A.D., Santiago D.I.,
Silbergleit A.S., 2002, Phys. Lett. A294, 79
\bibitem{Gon} Gonz\'{a}lez-D\'{\i}az P.F., 2000, Phys. Rev. D62, 023513
\bibitem{HE} Hawking S.W., Ellis G.F.R. {\it The Large Scale
Structure of Space-Time} Cambridge Univ. Press, Cambridge, UK,
1973
\bibitem{HKS} Hellman S., Kaloper N., Susskind L., 2001, JHEP 0106
\bibitem{Per} Perlmutter S., {\et al.}, 1997, \apj 483, 1565;
Riess A.G., {\it et al.}, 1998, Astron. J. 116, 1009


}
\end{iapbib}
\end{document}